\documentclass{article}

\usepackage[english]{babel}
\usepackage[utf8]{inputenc}

\usepackage{amsmath}
\usepackage{mathtools}
\usepackage{amsthm}

\usepackage{amsfonts}

\usepackage[scientific-notation=true]{siunitx}

\usepackage[dvipsnames,svgnames,x11names]{xcolor}
\usepackage[marginparwidth=1.7in]{geometry}
\usepackage[markup=underlined]{changes}

\usepackage[new]{old-arrows}
\usepackage{tikz-cd}

\definechangesauthor[color=BrickRed]{IR}
\definechangesauthor[color=NavyBlue]{MZ}

\usepackage{todonotes}
\setcommentmarkup{\todo[color={authorcolor!20},size=\scriptsize]{#3: #1}}

\newtheorem{theorem}{Theorem}[section]

\theoremstyle{remark}
\newtheorem{remark}[theorem]{Remark}
\theoremstyle{definition}
\newtheorem{definition}[theorem]{Definition}

\usepackage{float}
\usepackage{hyperref}

\title{Tensoring volatility calibration \\

\large
Calibration of the rough Bergomi volatility model via Chebyshev Tensors}

\author{Mariano Zeron \footnote{m.zeron@mocaxintelligence.com}   \\ Ignacio Ruiz\footnote{i.ruiz@mocaxintelligence.com}}
\date{August 2020}

\begin{document}

\maketitle

\begin{abstract}

Inspired by a series of remarkable papers in recent years (for example \cite{Andres}, \cite{bayer}, \cite{Horvath_vol_calibration}, \cite{Horvath_bayer_vol_calib}), that use Deep Neural Nets to substantially speed up the calibration of pricing models, we investigate the use of Chebyshev Tensors instead of Deep Neural Nets. Given that Chebyshev Tensors can be, under certain circumstances, more efficient than Deep Neural Nets at exploring the input space of the function to be approximated – due to their exponential convergence –  the problem of calibration of pricing models seems, a priori, a good case where Chebyshev Tensors can excel.
 
In this piece of research, we built Chebyshev Tensors --- either directly or with the help of the Tensor Extension Algorithms --- to tackle the computational bottleneck associated with the calibration of the rough Bergomi volatility model. Results are encouraging as the accuracy of model calibration via Chebyshev Tensors is similar to that when using Deep Neural Nets, but with building efforts that range between $5$ and $100$ times more efficient in the experiments run. Our tests indicate that when using Chebyshev Tensors, the calibration of the rough Bergomi volatility model is around $40,000$ times more efficient than if calibrated via “brute-force” (using the pricing function).

\end{abstract}

\section{Introduction}




OTC derivative pricing models are a central tool for financial institutions. For a given model, its price output depends on a number of parameters. Typically,  a range of them need to be specified before the pricing model can be used. Calibration is the specification of such parameters. The goal of calibration is to find the parameter values that allow the pricing model to replicate the different characteristics of the markets as best as possible.


Normally, derivative products can be priced by more than one pricing model. The choice of pricing model depends on different factors such as its mathematical tractability and questions of a practical nature.

When it comes to practical issues, a central consideration is the model's ability to describe market quotes; that is, how well it replicates market prices for a wide range of options.\footnote{It is market practice to quote prices as the implied volatilities linked to an option pricing model. The concepts of option price and implied volatility are so interlinked that both terms are often used interchangeably by market practitioners.} Another important characteristic is whether the model lends itself to calibration within the given constraints of the financial institution. For example, calibration is normally performed using optimisation routines. These often rely on running the pricing model a large number of times --- often thousands of times --- in the search for optimal parameters. If the pricing model is computationally expensive to run, calibration routines may not be executed within the time constraints of the business and would be of limited or no use in practice.

Unfortunately, mathematical tractability and practical use often pull in opposite directions; the better a model is at describing market quotes, the less mathematically tractable and slower it is to evaluate in a computer. As a result, relatively less powerful, but quicker and more tractable models are often chosen over financially strong but computationally expensive ones. This is sub-optimal.



\subsection{The challenge}

Since its presentation in $1973$, the Black-Scholes pricing model has been widely used for a variety of purposes. Part of its appeal is its mathematical tractability and simplicity. Over the years it became apparent that it lacked the power to describe market quotes; these typically display a range of elements that violate one of the main premises of the model: deterministic constant volatility.


In an attempt to correct the shortcomings of the Black-Scholes model, more powerful models have since been developed. Two widely used ones are the Heston (see \cite{Heston}) and the SABR model (see \cite{SABR}) --- along with their generalisations --- which model volatility in a stochastic manner. Although they capture some features of market implied volatilities, such as smile and skew, they still fail to replicate other aspects, such as the exploding power law nature of vol skew as time goes to zero (see \cite{Gatheral}). Despite some shortcomings, their mathematical tractability has made them popular among practitioners.


Rough volatility models lie at the opposite side of the spectrum to Black-Scholes. Introduced in \cite{rough_vol_first_paper}, these models have received a lot of attention in recent years. One of their main strengths is their ability to model some of the more complex intricacies displayed by market implied volatilities (see \cite{rough_bayer}, \cite{gatheral_rough_paper}). However, these models are less tractable, requiring slow Monte Carlo simulations for their use. This makes them, computationally speaking, very expensive to calibrate, creating bottlenecks in many applications that need to be done on a regular basis.


In recent years, the computational bottleneck of models --- such as the rough Bergomi model presented in \cite{rough_bayer} --- within pricing model calibrations has been addressed through the use of Deep Neural Nets (for example \cite{bayer}, \cite{Andres}, \cite{Horvath_vol_calibration}). In particular, in both \cite{bayer} and \cite{Horvath_vol_calibration}, the aim is to approximate the pricing model using a Deep Neural Net, which is then employed in the calibration exercise instead of the pricing model. 

This has several advantages. The training of the Deep Neural Net can be done offline, taking as much time as necessary. Once the training has been shown to give a good level of accuracy, the Deep Neural Net is used in the calibration exercise. This approach is computationally efficient as the evaluation of Deep Neural Nets amounts to little more than than simple linear algebra calculations. Moreover, the trained models can be serialised and used again in future calibrations.


In this paper, we use Chebyshev Tensors to approximate the pricing model. Inspired by the groundbreaking work in \cite{bayer}, \cite{Andres}, \cite{Horvath_vol_calibration}, the Chebyshev Tensor is built offline and used for the calibration exercises when needed. The aim is to show that with Chebyshev Tensors, pricing models such as rough Bergomi --- typically discarded for calibration, due to their computational cost --- can be used, within calibration routines, with minimal loss of accuracy. Moreover, for the test cases considered here, the results match (in some cases improve) those in \cite{bayer} and \cite{Horvath_vol_calibration}, in terms of accuracy and speed.

There are several key theoretical properties that justify the use of Chebyshev Tensors in this setting. The first is fast convergence for well-behaved functions: exponential for analytic; polynomial for smooth (\cite{TrefethenTextbook}). As the vast majority of pricing models behave (outside of isolated points) very well in terms of smoothness, the use of Chebyshev Tensors to approximate them makes sense. The second is that these tensors can be evaluated very efficiently and in a numerically stable manner (\cite{TrefethenTextbook}).

Their main drawback --- as is the case with any tensor --- is the curse of dimensionality: the number of grid points grows exponentially with the dimension of the tensor. The advantage in this context --- at least for the models considered here --- is that the dimension of the pricing functions approximated is not prohibitively large. Whenever possible, a single tensor is built. When the dimension is too high, we make use of Tensor Extension Algorithms, such as the ones presented in \cite{glau_completion}. Essentially, these algorithms extend the dimension for which a tensor can be used in a practical setting.



The paper is organised as follows. Section \ref{sec: model calibration} describes the framework we will use for pricing model calibration. Section \ref{sec: cheb tensors} gives a brief overview of Chebyshev Tensors; specifically what they are and which are their main properties. This same Section discusses the curse of dimensionality and how Tensor Extension Algorithms can be used to side-step it. Section \ref{sec: calibration with cheb tensors} describes how Chebyshev Tensors are built to speed up the calibration of pricing models. Section \ref{sec: numerical results} presents the numerical results obtained by applying Chebyshev Tensors to the calibration of the rough Bergomi model on synthetically generated data. The paper closes with a conclusion highlighting the main results and possible future research directions.

\section{Pricing Model Calibration}\label{sec: model calibration}







Let $\mathcal{D}$ be a derivative product type; for example an option. To obtain an instance of the derivative product, trade specific parameters must be specified. For example, maturity, payoff type, and strike determine an instance of an option $\mathcal{D}$.

The pricing function of the trade instance is a function $P$ that gives the price of the trade instance. We identify three sets of parameters, that $P$ depends on.

The first is the set of parameters $\Theta\subset\mathbf{R}^n$ of the model $\mathcal{M}$, that governs the dynamics of the derivative underlying. In the example mentioned above, where the trade type is an option, the model $\mathcal{M}$ governs the dynamics of the spot on which the option is defined. 
 The second set of parameters consists of $\Psi\subset\mathbf{R}^m$, which specify the instance of $\mathcal{D}$; for example time to maturity, strike and payment dates of the trade. The third family is $\Phi\subset\mathbf{R}^k$, which are the exogenous factors needed for pricing, such as spot values, interest rates, etc.

Once values $\theta\in\Theta$, $\psi\in\Psi$ and $\phi\in\Phi$, have been specified, the function

\begin{equation}\label{eq: pricing function in terms of model}
P(\mathcal{M(\theta), \psi, \phi})
\end{equation}

\noindent returns a price for the specific instance of $\mathcal{D}$.


For a given trade type, market quotes come in a range of maturities and strikes. These quotes appear in the market as the monetary value (i.e. price) of the trade but are often expressed in terms of implied volatilities. Implied volatilities are obtained by inverting a pricing model, so that given price, maturity, strike and other exogenous factors (the ones in $\Phi$), an implied volatility is obtained. From now on, we assume the Black-Scholes pricing model is used for this purpose --- it is the one typically used in the industry and the one used to produce the results presented in Section \ref{sec: numerical results}. The collection of implied volatilities (obtained by varying maturities and strikes) for a given trade type is commonly referred to as the \emph{implied volatility surface}. \footnote{Given this relation between prices and implied volatilities, practitioners often make no distinction between them.}

As explained above, a pricing function (Equation \ref{eq: pricing function in terms of model}) returns prices; that is, the monetary value of a trade. To compare the output of a pricing function with market quotes, we turn the prices returned by Equation \ref{eq: pricing function in terms of model} into implied volatilities using the Black-Scholes pricing model. This process is expressed by the following composition of functions

\begingroup
\large
\begin{equation}\label{eq: implied vol function price followed by bs}
\begin{tikzcd}
\varphi(\mathcal{M}(\theta), \psi, \phi):\mathbb{R}^{n+m+k}\rar{P} & \mathbb{R}^n \rar{P^{-1}_{BS}}  & \mathbb{R}
\end{tikzcd} 
\end{equation}
\endgroup
where $P$ is the pricing function as in Equation \ref{eq: pricing function in terms of model}, that depends on the three families of parameters $\Theta, \Psi, \Phi$; and $P_{BS}$ is the Black-Scholes pricing model used to invert prices (monetary value of trade) to implied volatilities. From now on, we will also refer to $\varphi$ as a pricing function.


Calibration is the exercise through which a set of values in $\Theta$ is specified, so that that the implied volatilities obtained through $\varphi$ (Equation \ref{eq: implied vol function price followed by bs}), are matched as closely as possible to the market quotes.


Now, for a given $\theta$, the implied volatilities produced by the pricing model may or may not be close to the market quotes. How close they are is typically measured through a loss or cost function. The following is an example of a loss function

\begin{equation}\label{eq: Loss function}
    L(\theta, \phi) = \sum^{N}_{i = 1} \omega_i(q_i - v_i(\theta, \phi))^2,
\end{equation}

\noindent where there are $N$ market quotes that constitute the implied volatility surface --- different maturities and strikes contained in $\Psi$ --- $q_i$ is the $i$-th implied volatility obtained from market quotes, $v_i$ the $i$-th implied volatility obtained with the pricing model (Equation \ref{eq: implied vol function price followed by bs}), and $\{w_i\}$ a set of weights.

In terms of the loss function, calibration consists of finding a set of values in $\Theta$ that minimise it. That is

\begin{equation}
    \theta^{*} = \underset{\theta\in\Theta\subset\mathbf{R}^n}{argmin} \ \   L(\theta, \phi)
\end{equation}


The minimisation of the loss function rarely has an analytic solution, therefore optimisation routines must be employed. These generally consist of trial and error exercises; at each trial an implied volatility surface is produced by Equation \ref{eq: implied vol function price followed by bs} and compared to market quotes. Typically, optimisation routines involve hundreds if not thousands of these trials. Therefore, the pricing model is called thousands if not tens of thousands of times.


Different types of optimisation routines can be used. There are optimisation routines that use the gradient of the function to minimise; these tend to be faster but do not guarantee a global minimum. Popular examples of these are SLSQP, L-BFGS-B and Levenberg-Marquardt. On the other hand, there are gradient-free optimisers; these  have gathered a lot of attention in recent times. They tend to be slow but guarantee a global solution. Moreover, they are able to deal with a mix of variables (discrete and continuous) and large dimensions. Popular examples of these are COBYLA, Nealder-Mead and genetic algorithms. The choice of which routine to use in a given context depends on a number of factors, such as the loss function used and the computational time that it takes to run.


This paper gives evidence that Chebyshev Tensors allow for timely and accurate calibration of expensive models such as the rough Bergomi (see \cite{rough_bayer}). This is done by approximating $\varphi$ (Equation \ref{eq: implied vol function price followed by bs}), restricted to the parameter set $\Theta$ (space over which calibration is done), with a Chebyshev Tensor, and using this Chebyshev Tensor instead of $\phi$ at calibration.

The intention is not to examine a range of optimisation algorithms that can be used; we will restrict our attention to the ones that serve our needs. Details of this will be given in Sections \ref{sec: calibration with cheb tensors} and \ref{sec: numerical results}.

\subsection{Rough Bergomi model}\label{subsec: rough bergomi}

The following is the pricing model used in the tests for which numerical results are shown in Section \ref{sec: numerical results}.

Using the notation in Section \ref{sec: model calibration}, the rough Bergomi model $\mathcal{M}(\Theta)$, has parameter space $\Theta = (\widetilde{\xi}, \eta, \rho, H)$. The stochastic differential equation that governs the dynamics of the asset is 
\begin{align*}
    dS_t &= -0.5V_tdt + \sqrt{V_t}dW_t, \ \ \ \ \textrm{for}\ t > 0,  \ \ \ \ S_0 = 0,\\
    V_t &= \widetilde{\xi}(t)\varepsilon\Big( \sqrt{2H}\eta \int_0 ^t (t-s)^{H - 0.5} dZ_s \Big) \ \ \ \ \textrm{for}\ t > 0,  \ \ \ \ V_0 > 0,
\end{align*}
where $\widetilde{\xi}$ is the initial forward variance curve (see \cite{bergomi} Section $6$), $\varepsilon$ the stochastic exponential (\cite{Doleans}), $\eta>0$ the volatility of $V_t$, $H$ the Hurst parameter in $(0,1)$, and $\rho$ a correlation parameter in $[-1,1]$ that relates the two stochastic drivers $Z$ and $W$.

For numerical calculations, the curve $\widetilde{\xi}$ is approximated by a piecewise constant term structure. Along with the rest of the parameters $\eta$, $\rho$ and $H$, these constitute $\Theta$ in $\mathbb{R}^n$, for some $n\in\mathbb{N}$. The numerical tests presented in Section \ref{sec: numerical results} considered the case where $\widetilde{\xi}$ is constant --- giving a parameter space $\Theta$ of dimension $4$ --- and the case where $\widetilde{\xi}$ is piecewise constant with $8$ points --- defining $\Theta$ in dimension $11$.

For general details on rough volatility models and their numerical simulations, we refer the reader to \cite{Horvath_rough_vol}.

\section{Chebyshev Tensors}\label{sec: cheb tensors}



First we cover the main properties of Chebyshev Tensors as functions approximators. Then we discuss the curse of dimensionality --- one of the main challenges when using tensors --- and see how it can be side-stepped using Tensor Extension Algorithms.

\subsection{Main properties}

A \emph{tensor} is simply a collection of points in Euclidean space --- usually referred to as grid points or mesh --- with a real value associated to each point. The distribution of grid points can in principle be any. In what follows, the grids considered are built by taking the Cartesian product of grids defined in each dimension.

\begin{figure}[H]
\centering
\includegraphics[scale=0.5]{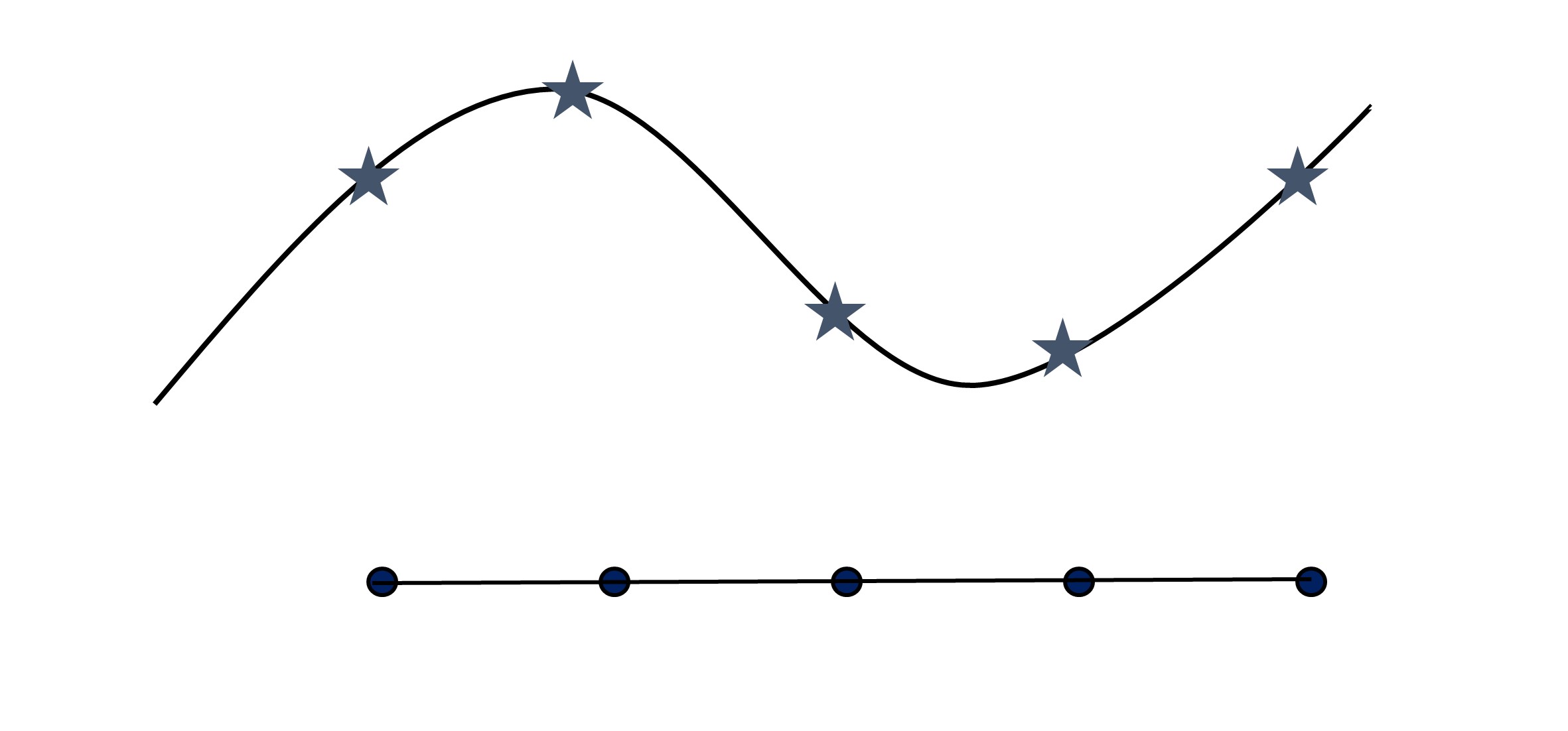}
\caption{Tensor defined by evaluating a function on a grid of points on the function's domain.}
\label{fig: tensor from function}
\end{figure}

We use tensors to approximate real functions. An illustration in dimension one shows how. Let $f$ be a one-dimensional function with bounded domain $[a,b]$. Given any grid of points on $[a,b]$, the function $f$ evaluates every grid point (as in Figure \ref{fig: tensor from function}) to obtain a tensor $T$ associated to $f$.\footnote{The \emph{dimensionality} of the function is the dimension of the domain where the function takes values. The image of the function will always be of dimension one, unless otherwise stated.}

There are different ways in which the tensor $T$ can be extended to a function over the domain of $f$. We know that given a tensor with $n$ points, there is a unique polynomial $p$, of degree less than or equal to $n-1$, that interpolates the values of the tensor at the tensor points. This polynomial can be used as a proxy for $f$ on its domain $[a,b]$. If the function $f$ is difficult to compute, having a $p$ that approximates $f$ to a high degree of accuracy could prove useful in a range of applications, as $p$ is typically easy to evaluate. The natural question is if we can ensure $p$ to be an accurate proxy to $f$. From now on, we use tensors and their corresponding polynomial interpolants interchangeably.

Intuition says that the more grid points, the more information collected from $f$ and hence the higher the accuracy. However, this is not the case, as the Runge phenomenon shows (\cite{Runge}). The Runge function is a one-dimensional analytic function; in particular it is infinitely differentiable. However, the interpolants $p$ associated to tensors defined on equidistant grids --- in some sense a natural choice --- diverge away from $f$ exponentially as the number of grid points increases. 

The choice of grid is fundamental to control the problem exhibited by the Runge phenomenon. As we will see, this type of behaviour disappears when the correct geometry of the grid is used and applied to the right class of functions.

There are different families of grid points that correct the problems of the Runge phenomenon. The choice in this paper, for reasons explained in \ref{rmk: why cheb and not others}, are \emph{Chebyshev points}.

\begin{definition}
The Chebyshev points associated with the natural number $n$ are the real part of the points

\end{definition}
\begin{equation}\label{dfn: cheb points}
x_j = \mathrm{Re}(z_j) = \frac{1}{2}(z_j+z_j^{-1} ),\ \ \ \ \ 0\leq j \leq n.
\end{equation}

Equivalently, Chebyshev points can be defined as

\begin{equation}
x_j = \mathrm{cos} \Big( \frac{j\pi}{n} \Big), \ \ \ \ \ \ 0\leq j \leq n.  
\end{equation}

\begin{figure}[H]
\centering
\includegraphics[scale=0.5]{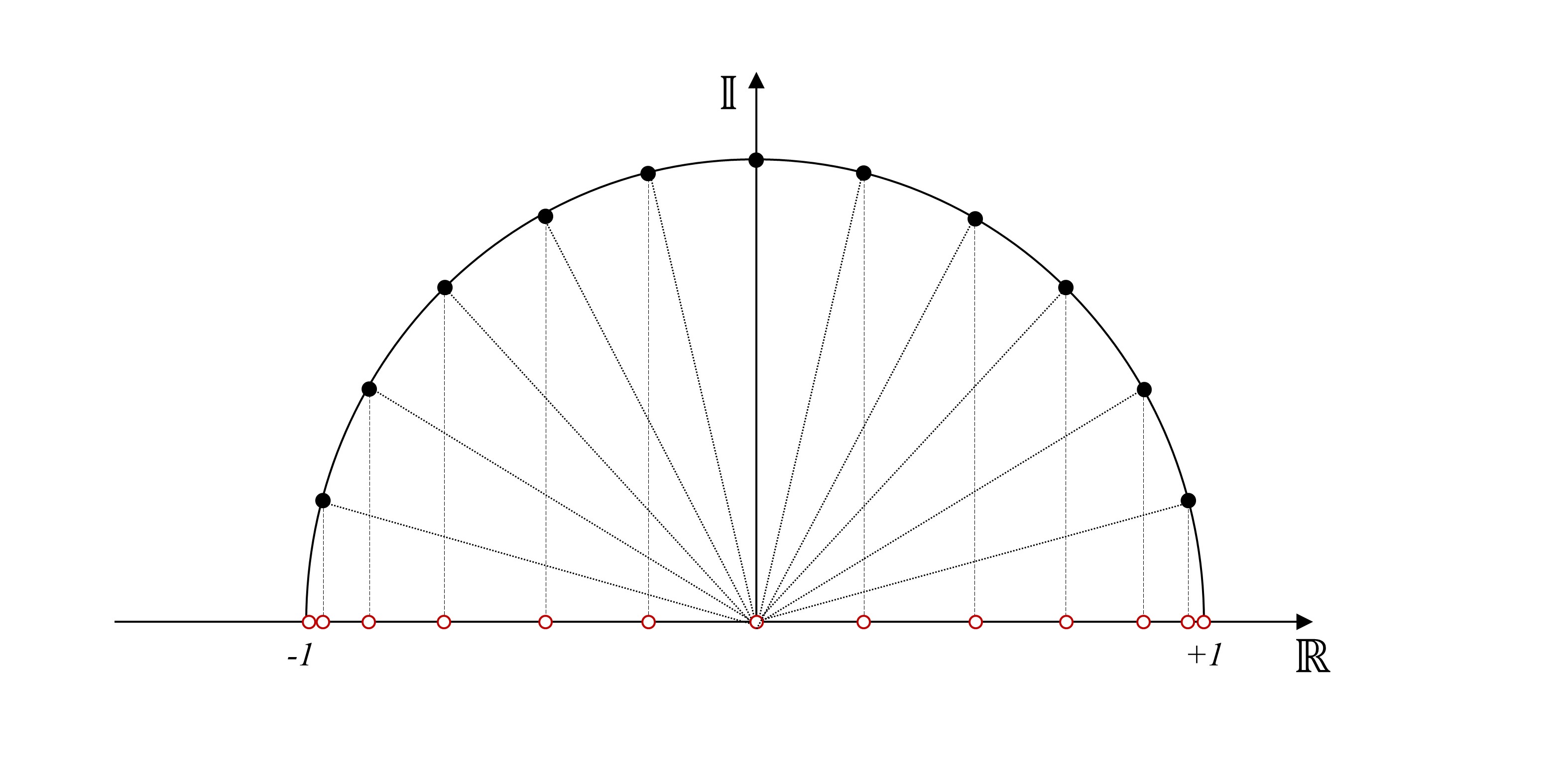}
\caption{Chebyshev points in one dimension.}
\label{fig: 1d cheb mesh}
\end{figure}

As can be seen in Figure \ref{fig: 1d cheb mesh}, Chebyshev points are the projection onto the real line of equidistant points on the upper-half of the unitary circle.

The definition of Chebyshev points is easily extended to arbitrary closed and bounded intervals $[a,b]$ using linear mappings and translations. From now on, everything stated for the interval $[-1,1]$ is equally valid for $[a,b]$.

The extension of Chebyshev points to higher dimensions is obtained by taking Cartesian products of Chebyshev points in dimension one.

\emph{Chebyshev Tensors} are tensors that have Chebyshev points as grid points. Their corresponding polynomial interpolants are \emph{Chebyshev interpolants}. One of the main strengths of tensors defined over these points is the convergence rates they exhibit. If the function $f$ to be approximated is Lipschitz continuous --- a mild restriction on continuity --- Chebyshev Tensors converge to $f$ as the number of grid points increases; if $f$ is smooth, convergence is polynomial; when $f$ is analytic, convergence is exponential in dimension one and quasi-exponential in higher dimensions. The following Theorem states the latter. The proof can be found in \cite{GlauParamOptPric}.


\begin{theorem}\label{thm: exponential multidim}
Let $f$ be a $d$-dimensional analytic function defined on $[-1,1]^d$. Consider its analytical continuation to a generalised Bernstein ellipse $E_p$, where it satisfies $\|f\|_{\infty}  \leq M$, for some $M$. Then, there exists a constant $C>0$, such that

\begin{equation}
\|f-p_n \|_{\infty}\leq C\rho^{-m} 
\end{equation} 

\noindent where $\rho=min_{(1\leq i\leq d)} \rho_i$, and $m=min_{(1\leq i\leq d)} m_i$. The collection of values $\rho_i$ define the radius of the generalised Bernstein ellipse $E_p$, and the values $m_i$ define the size of the Chebyshev mesh.
\end{theorem}

For more details on Theorem \ref{thm: exponential multidim}, its proof and related results, see \cite{GlauParamOptPric}. The one-dimensional version, which shows exponential convergence, can be found in \cite{TrefethenTextbook}.

One of the advantages of working with Chebyshev Tensors is that the evaluation of their corresponding polynomial interpolants can be done in a numerically stable and fast way through the use of the Barycentric Interpolation Formula or the Clenshaw Algorithm. This is essential in practical settings. If numerically unstable algorithms are used to evaluate these polynomials --- such as the Vandermode algorithm, which is exponentially unstable (chapter $5$ of \cite{TrefethenTextbook}) and often implemented in popular software packages such as SciPy --- one can get unexpected errors in the results.

On top of being numerically stable and fast, the Barycentric Interpolation Formula and the Clenshaw Algorithm only need the information at the level of the tensor --- that is, the distribution of grid points and the values associated to them. For further details, we refer to \cite{TrefethenTextbook} regarding the Barycentric Interpolation Formula and to \cite{Clenshaw_stability} for the Clenshaw Algorithm.


Another important property of Chebyshev Tensors is their ability to approximate the derivatives of $f$. Given that the proxies to $f$ obtained from Chebyshev Tensors are polynomial interpolants, it is not surprising that their derivatives are easy to obtain. What is surprising is that their derivatives are good proxies to the derivatives of $f$. The following Theorem formally expresses this for dimensions greater than one. It states that the partial derivatives of the Chebyshev Tensors $p_n$ to an analytic function $f$ converge polynomially to the partials of $f$. Its proof along with further details can be found in \cite{GlauParamOptPric}.

\begin{theorem}\label{thm: polynomial convergence derivatives hd}
Let $f$ be a $d$-dimensional analytic function defined on $\mathcal{A} = [-1,1]^d$. Consider its analytic continuation to a generalised Bernstein ellipse $E_p$. Then for every $m\in \mathbb{N}$, there exists a constant $K(m)$ such that
\begin{equation*}
\| f - p_n \|_{C^l(\mathcal{A})} \leq K(m)n^{-m}\|f\|_{C^{2(l+1) + d + m}(\mathcal{A})}
\end{equation*}
where
\begin{equation*}
\| f \|_{C^l(\mathcal{A})} = \underset{|\alpha |\leq l}{\textrm{max}} \ \underset{x \in \mathcal{A}}{\textrm{max}}\  |\partial^{\alpha}f(x) |.
\end{equation*}
\end{theorem}


When the function $f$ has dimension one, the result can be strengthened to exponential convergence. For details of this case we refer to \cite{TrefethenTextbook}.


\bigskip

\begin{remark}\label{rmk: why cheb and not others}
Chebyshev points are not the only family of points that have properties such as the ones illustrated in Theorem \ref{thm: exponential multidim}. Other examples such as Legendre, Jacobi and Gegenbauer, along with their corresponding Tensors and interpolants, enjoy similar properties. However, our preference is always Chebyshev. Out of all these families, Chebyshev Tensors are by far the simplest to generate and evaluate; paramount characteristics in practical applications such as the ones presented in this paper.
\end{remark}

\subsection{Curse of dimensionality}



The curse of dimensionality is a problem that afflicts tensors of any kind. It states that as the dimension increases, the number of points on the grid increases exponentially. This presents a problem in practical settings as one quickly reaches a dimension where it is no longer feasible to store the grid in memory nor evaluate the function on all grid points.

There are many applications where the dimension is greater than what a single tensor can manage. For these cases, techniques to side-step the curse of dimensionality must be explored. There are different techniques which may be considered; the choice partly depends on the application. For example, in \cite{cheb_slider}, tensors are built through the use of Orthogonal Chebyshev Sliders to approximate functions with dimension in the hundreds. However, for the problem concerning this paper, and given the dimension of the functions in question --- always less than $20$ --- the Tensor Extension Algorithms presented in \cite{glau_completion} are more appropriate.


The Tensor Extension Algorithms presented in \cite{glau_completion} require working with tensors expressed in TT format. The following describes the main properties needed.

\subsubsection{Tensors in TT format}

Consider a grid of dimension $d$, with $n_i$ points per dimension, where $1 \leq i \leq d$. The grid has a total of $n_1 \cdot n_2 \cdots n_d$ points. To define a tensor $\mathcal{X}$ on this grid, we only need to assign real values on each of the grid points.

Define $d$ vectors $C_i$, each with $n_i$ real values. We define the following tensor $\mathcal{X}$ using the vectors $C_i$ in the following manner

\begin{equation}\label{eq: tt tensor rank 1}
\mathcal{X}(j_1, \ldots, j_d) = C_1(j_1)\cdots C_d(j_d).
\end{equation}

\noindent where $j_i$ is an index such that $1\leq j_i \leq n_i$, for all $i$, $1\leq i\leq d$.

Notice that the tensor $\mathcal{X}$ defined in Equation \ref{eq: tt tensor rank 1} is an $d$-dimensional tensor with $n_1 \cdot n_2 \cdots n_i$ points. However, the tensor is fully determined by $n_1 + n_2 + \cdots + n_d$ values; the ones in the vectors $C_i$.

To make things concrete, consider the following example. let $d=2$, $n_1=4$ and $n_2=3$. Let $C_1=(1.6, 2.1, -3.2, 8.4)$ and $C_2=(7.4, -6.1, 9,5)$. Then, $\mathcal{X}(2,3)= 2.1 \times 9.5 = 19.95 $.

The example in Equation \ref{eq: tt tensor rank 1} is generalised as follows. Instead of vectors of real values, $C_i$ can be a vector of matrices, all of the same rank. This means that $C_i(j_i)$ is a matrix of a pre-specified rank, for all $i$, $1\leq i\leq d$, and all $j_i$, $1\leq j_i\leq n_i$. The definition can be formalised as follows

\begin{definition}\label{dfn: tensors in TT format}
Let $\mathcal{X}$ be a tensor of dimension $d$ and let its mesh be defined by $n_i$ points in $i$-th dimension for all $i$, $1\leq i\leq d$. The tensor $\mathcal{X}$ can be expressed in TT format with ranks $(r_0, \ldots, r_{d})$, if the following hold,
\begin{enumerate}
    \item There is a set of vectors of matrices $C_i$, called the \emph{cores} of $\mathcal{X}$, where $1\leq i\leq d$.
    \item $C_i(j)$ is a matrix of rank $r_{i-1}\times r_{i}$, for $1\leq j \leq n_i$.
    \item The first and last rank are one; that is $r_0 = r_d = 1$.
    \item The values of $\mathcal{X}$ on the grid can be recovered as follows,
\end{enumerate}
\begin{equation}\label{eq: tensor TT rank r}
\mathcal{X}(j_1, \ldots, j_d) = C_1(j_1)\cdots C_d(j_d).
\end{equation}
\end{definition}

Let us illustrate with the following example. Let $d=3$, $n_1=4$,  $n_2=3$, $n_3=2$. Consider the following matrices
\begin{equation*}
C_1=\left(\begin{bmatrix}1.3 & -2.8\end{bmatrix}, \begin{bmatrix}9.7 & 4.8\end{bmatrix}, \begin{bmatrix}-2.4 & 6.9\end{bmatrix}, \begin{bmatrix}8.5 & -2.1\end{bmatrix}\right),  
\end{equation*}

\begin{equation*}
C_2=\left(\begin{bmatrix}9.7 & -9.5 & -7.5\\4.9 & -9.2 & 3.8\end{bmatrix}, \begin{bmatrix}4.8 & 8.2 & 6.5\\-8.9 & -2.6 & -8.3\end{bmatrix}, \begin{bmatrix}4.2 & -2.7 & -4.9\\1.9 & 2.2 & 1.3\end{bmatrix}\right),
\end{equation*}

\noindent and 
\begin{equation*}
C_3=\left(\begin{bmatrix}-3.7 \\ -2.5 \\ 7.9\end{bmatrix}, \begin{bmatrix}2.5 \\ 6.8 \\ -5.4\end{bmatrix}\right). 
\end{equation*}

Then, $\mathcal{X}(2,3,2)= \begin{bmatrix}9.7 & 4.8\end{bmatrix} \cdot \begin{bmatrix}4.2 & -2.7 & -4.9\\1.9 & 2.2 & 1.3\end{bmatrix} \cdot \begin{bmatrix}2.5 \\ 6.8 \\ -5.4\end{bmatrix}$.

The main advantage of working with tensors in TT format is the huge reduction in storage cost. The cost of $\mathcal{O}(n^d)$ associated to an ordinary tensor is reduced down to $\mathcal{O}(dnr^2)$, where $r$ is the maximum of the ranks in $r_i$. If $r$ is low, then the reduction in cost as the dimension grows is considerable --- the exponential growth in terms of the dimension $d$ is changed to linear.

The second advantage concerns the evaluation of these tensors. This is defined via the inner product of tensors which can be defined in the following way

\begin{equation}\label{eq: tensor inner product}
\langle \mathcal{X}_1, \mathcal{X}_2 \rangle(i_1, \ldots, i_d) = \sum_{i_1 = 1}^{n_1} \cdots \sum_{i_d = 1}^{n_d}\mathcal{X}_1(i_1, \ldots, i_d)\mathcal{X}_2(i_1, \ldots, i_d).
\end{equation}

We will not go into the details of how the inner product is used to evaluate tensors in TT format. This is beyond the scope of the paper. For details on this and general results on tensors in TT format, we refer the reader to \cite{glau_completion} and \cite{TT_tensor_Oseledets}. However, building tensors in TT format and evaluating them lies at the core of the application in this paper.


\subsubsection{Tensor Extension Algorithms}\label{subsec: tensor extension}

In this Section we describe how the Tensor Extension Algorithms in \cite{glau_completion} can be used to obtain some of the results presented in Section \ref{sec: numerical results}.

A given tensor $\mathcal{T}$ of dimension $d$ has a building and storage cost of $\mathcal{O}(n^d)$. The corresponding cost of a tensor in TT format $\mathcal{X}$ is $\mathcal{O}(dnr^2)$. If the rank $r$ is fixed and $d$ increases, the difference between these costs becomes considerable. If a tensor $\mathcal{X}$ in TT format approximates $\mathcal{T}$ to a high degree of accuracy, it makes sense to use it instead of $\mathcal{T}$. The Tensor Extension Algorithms in \cite{glau_completion} find such tensors $\mathcal{X}$.


There are three main Tensor Extension Algorithms in \cite{glau_completion}. The first and most fundamental is called the \emph{Completion Algorithm}. This considers the space $\mathcal{F}_{r}$ of tensors in TT format of a fixed rank $r = (r_0, r_1, \ldots, r_d)$. One of the main properties of this space is that it is a smooth Riemannian manifold. This means it can be navigated without encountering sharp edges. 

The Completion Algorithm does the following on $\mathcal{F}_{r}$. Let $\mathcal{T}$ be a tensor to be approximated by some tensor $\mathcal{X}$ in $\mathcal{F}_{r}$. Assume it is inconvenient (perhaps impossible) to evaluate all grid points of $\mathcal{T}$; otherwise this would be done. Therefore, only a fraction of the grid is evaluated. This is all the information used to find $\mathcal{X}$. The aim is to minimise the distance between $\mathcal{T}$ and the space $\mathcal{F}_{r}$, restricted to the information available; that is, the set of grid points $\mathcal{K}$ at which $\mathcal{T}$ was evaluated. Specifically, we want the following

\begin{equation}\label{eq: completion algo loss}
\underset{\mathcal{X}\in \mathcal{F}_r}{min} \|\mathcal{T}_{\mathcal{K}} - \mathcal{X}_{\mathcal{K}}\|^2 
\end{equation}

\noindent where $\mathcal{X}_\mathcal{K}$ means $\mathcal{X}$ restricted to the subgrid $\mathcal{K}$.

As the space $\mathcal{F}_{r}$ is smooth, the minimisation can be done efficiently using optimisation algorithms such as the Conjugate Gradient method. As is done in most optimisation exercises --- for example, when training a Deep Neural Net --- the data set $\mathcal{K}$ is split into a training and a testing set. The choice of what tensor $\mathcal{X}$ to try at each iteration is driven by the loss on the training set; the verification of whether the algorithm has converged is done using both training loss and testing loss.

The Completion Algorithm described above restricts its search to a space $\mathcal{F}_r$ with fixed rank. The \emph{Rank Adaptive Algorithm} enhances the Completion Algorithm by increasing the rank of the space over which the search is performed. This is done in the following way.

Start with the space $\mathcal{F}_r$, where $r = (r_0, r_1, \ldots, r_d)$. If the Completion Algorithm is run on this space and no tensor $\mathcal{X}$ with the desired degree of accuracy is found, then the Rank Adaptive Algorithm increases the value of one of the ranks $r_i$ in $r$, where $i$ can be any value between $1$ and $d-1$. Intuitively, the idea is to increase the size of $\mathcal{F}_r$, thereby increasing the chances of finding a tensor $\mathcal{X}$ with the right accuracy.

Normally, the Rank Adaptive Algorithm starts by running the Completion Algorithm on $\mathcal{F}_r$, where $r_i = 1$, for all $i$, $0\leq i\leq d$. If no suitable $\mathcal{X}$ is found for this rank, $r_1$ is increased by $1$. The new search performed by the Completion Algorithm is now done on $\mathcal{F}_r$, where $r_1 = 2$ and $r_i = 1$, for all $i\neq 1$.

This process is repeated, for every $r_i$, going from $i = 1$, to $i = d-1$, and then starting again with $i = 1$. This continues until a suitable $\mathcal{X}$ is found, or the algorithm is forced to stop --- for example, after a pre-determined number of iterations.

The last of the three algorithms presented in \cite{glau_completion} is the \emph{Sample Adaptive Algorithm}. This consists of increasing the size of $\mathcal{K}$, from Equation \ref{eq: completion algo loss}, every time the Rank Adaptive Algorithm does not return a suitable $\mathcal{X}$. 


The algorithms mentioned so far give no guarantee of convergence. That is, given a tensor $\mathcal{T}$, there is no guarantee that a tensor $\mathcal{X}$ in TT format that approximates $\mathcal{T}$ to any given degree of accuracy is found. In practice, the evidence points towards positive results when $\mathcal{T}$ is reasonably well behaved (for example \cite{glau_completion} and \cite{Completion_Steinlechner}). The results presented in section \ref{sec: numerical results} are further empirical evidence that for some pricing functions in finance, these algorithms do indeed find suitable tensors $\mathcal{X}$ to work with, side-stepping the curse of dimensionality.

\section{Calibration with Chebyshev Tensors}\label{sec: calibration with cheb tensors}


In Section \ref{sec: model calibration}, the calibration of a pricing model was defined as the minimisation of the loss functions in Equation \ref{eq: Loss function}, where values $v_i(\theta, \phi)$ are obtained by calling the pricing function and then computing their corresponding implied volatilities. The minimisation of Equation \ref{eq: Loss function} is normally done using optimisation algorithms. These rely on computing implied volatility surfaces --- that is the values $v_i(\theta, \phi)$ --- for every parameter point in the domain explored by the algorithm. If the pricing function is expensive to run --- such as is the case for the rough Bergomi model --- then the optimisation exercise can be prohibitively expensive.


The idea will be to approximate the pricing function defined in Equation \ref{eq: implied vol function price followed by bs}, with Chebyshev Tensors, and use the latter in the calibration process.
As Chebyshev Tensors are very fast to evaluate, the calibration exercise can be done in less than a second (see results in Section \ref{sec: numerical results}). If the accuracy of approximation of the pricing model is high --- as is the case for the results presented in Section \ref{sec: numerical results} --- one can be confident that the calibration obtained using the Chebyshev Tensor is essentially as good as the one obtained using the pricing model.


Another advantage of using Chebyshev Tensors as replacements of pricing models in calibration exercises is the following. The optimisation algorithms employed sometimes rely on the gradient of Equation \ref{eq: Loss function}. A key property of Chebyshev Tensors is that their partial derivatives --- and hence their gradient --- are easy to compute and accurate (Theorem \ref{thm: polynomial convergence derivatives hd}). This means the gradient of Equation \ref{eq: Loss function} can be obtained very easily and used in a wide range of optimisation algorithms.


There are two ways in which Chebyshev Tensors are built for calibration exercises.\footnote{These are the ones so far considered by the authors and for which results are shown. Others may exist.} Both allow for the tensor to be built offline. Once the tensor has been built and its accuracy verified by direct comparison with the pricing model, it can be used in calibration exercises as many times as needed.


A big advantage of splitting the process in building (offline) and evaluating (online) is that existing calibration systems do not need to be modified. Chebyshev Tensors become an added tool for the practitioner to test pricing models and run fast calibrations to market quotes.


The first step in the building process is to define the Chebyshev grid over the domain of approximation. For the tests done, this consisted of the parameter space $\Theta$ along with maturity and strike. Note that Chebyshev Tensors could have been built for $\Theta$ alone, fixing maturity and strike. However, this would require building a Chebyshev Tensor for each of the implied volatilities needed. They can be also built for $\Theta$ and strike, fixing maturity (or $\Theta$ and maturity, fixing strike). This would give Chebyshev Tensors that approximate the smile of the vol surface for a fixed maturity. Again, as many Chebyshev Tensors as maturities needed for calibration would have to be built. Despite the added cost incurred by adding maturity and strike to the domain of the tensors, we opted to include them, as this would yield tensors capable of giving implied volatilities at any point of the vol surface, which is the optimal setup.

\subsection{Defining the tensor}


To define the grid, ranges for each of the parameters that constitute the space $\Theta$, along with ranges for maturity and strike, must be specified. Ideally, these ranges should be large enough so that a good calibration can be obtained. Once the ranges have been defined, one must choose the number of Chebyshev points for each parameter. This is all the information needed to determine the grid of the tensor.


Regarding the real values, one possible approach at this point is to evaluate the pricing model at each grid point. This is possible when the dimension of the domain is roughly between $1$ and $6$. Dimensions greater than this produce grids which are too large to handle due to the curse of dimensionality. If a tensor $\mathcal{T}$ is built this way --- which we call the ``direct'' way --- the building process finishes here. This is the case, for example, of the rough Bergomi model with constant forward variance for which results are presented in Section \ref{sec: numerical results}.


When the dimension of $\Theta$ is too large to build $\mathcal{T}$ as described above, one must use an alternative, cheaper way of building the tensor. In this paper, we opt for Tensor Extensions Algorithms as presented in \cite{glau_completion}, described in Section \ref{subsec: tensor extension}.


As explained in Section \ref{subsec: tensor extension}, first a random selection of Chebyshev points is made. The implied volatilities for these grid points are then computed. This information is used by the Tensor Extension algorithms --- in our case the Rank Adaptive Algorithm --- to produce a tensor in TT format $\mathcal{X}$ that is a proxy to $\mathcal{T}$. The tensor $\mathcal{X}$ is stored and used in calibration exercises.


As Theorem \ref{thm: exponential multidim} indicates, Chebyshev Tensors are very good at approximating functions within a given domain. However, outside of it, there is no guarantee. This is an important element to bear in mind when using Chebyshev Tensors in calibration exercises due to the following.


The optimisation algorithms used to minimise Equation \ref{eq: Loss function} rely on sampling the parameter domain $\Theta$. If a Chebyshev Tensor is used instead of the pricing function in the calibration exercise, it will only give accurate implied volatilities if the parameter point explored by the optimisation algorithm is within the domain of approximation. Therefore, it is preferable to use optimisation algorithms that admit domain restrictions. This ensures the parameter points evaluated by the Chebyshev Tensor are within the domain where we know the accuracy to be high. There are plenty of optimisation algorithms that admit domain restrictions; for example SLSQP, L-BFGS-B and Least Squares.


If an optimisation algorithm that does not admit domain restrictions must be used, one can opt for a mix of Chebyshev Tensor and pricing function calls. In this case, Chebyshev Tensors return implied volatilities only when the parameter explored by the optimisation algorithm is within the domain of approximation. When it is not, the pricing function is used. This should be an effective approach as the domain of Chebyshev Tensors can be chosen large enough to reduce the chances of having to evaluate parameters outside this domain. One can, for example, use parameters from historical calibrations to identify suitable domains. Notice that increasing the domain of the Chebyshev Tensors should barely affect their accuracy given the quasi-exponential convergence presented in Theorem \ref{thm: exponential multidim}.

\section{Numerical Results}\label{sec: numerical results}


The pricing model used to generate the results presented in this Section is the rough Bergomi model (see Section \ref{subsec: rough bergomi}). This was used to compute European call option prices. Two variants were considered based on the number of dimensions used for the forward variance parameter $\widetilde{\xi}$. The first set of tests assumed constant forward variance. The second considered a term structure with $8$ points.


The implementation of the pricing routine used is based on the one mentioned in \cite{bayer}, available on github and implemented in C++.\footnote{The code can be found in https://github.com/roughstochvol} It is a Monte Carlo based pricing routine; $60,000$ paths were used to generate the prices in our tests. Once prices are obtained, these are turned into implied volatilities.\footnote{Implied volatilities are also obtained using functions in https://github.com/roughstochvol, and relies on the Newton-Ralphson method implemented in boost.}.


Chebyshev Tensors were built in two different ways. First, in a ``direct" way. This consists in evaluating every grid point. We call Chebyshev Tensors built this way \emph{Direct Chebyshev Tensors}. When the forward variance was constant, the dimension of the tensor is $6$ ($4$ from the parameters $\Theta$, $2$ from maturities and strikes). Therefore, it was possible to build Direct Chebyshev Tensors. In this case, Chebyshev Tensors were also built using the Rank Adaptive Algorithm to evaluate which of the two methods yields better results. When the forward variance was modelled using $8$ points, the tensor had dimension $13$ ($11$ from $\Theta$, $2$ from maturities and strikes). Therefore, Chebyshev Tensors were only built using the Rank Adaptive Algorithm.


Direct Chebyshev Tensors were built in python using the MoCaX library.\footnote{Available at https://www.mocaxintelligence.com/download-mocax-intelligence/} Chebyshev Tensors built using the Rank Adaptive Algorithm used the Matlab implementation described in \cite{Completion_Steinlechner}.


A total of $1,000$ implied volatility surfaces were used for calibration. These were synthetically generated by the rough Bergomi pricing model. Each surface is generated by specifying $1,000$ parameter combinations and the following maturities and strikes.\footnote{Maturities are in years. The underlying for the European option is assumed to have value $1$.}

\begin{itemize}
    \item     \textrm{maturities} = \{ 0.3, 0.6, 0.9, 1.2, 1.5, 1.8, 2.0 \}
    \item     \textrm{strikes} = \{ 0.7, 0.75, 0.8, 0.85, 0.9, 0.95, 1.0, 1.05, 1.1, 1.15, 1.2, 1.25, 1.30 \}
\end{itemize}

The $1,000$ parameter combinations were obtained by randomly sampling (assuming a uniform distribution) the parameter domain of the rough Bergomi model. The domain over which the parameters were sampled is given by the following

\begin{equation*}
    (\widetilde{\xi}, \eta, \rho, H) \in [0.01, 0.16]\times [0.5, 4]\times [-0.95, -0.1]\times [0.025, 0.5].
\end{equation*}




The ranges above are the same as the ones chosen to train Deep Neural Nets in \cite{Horvath_vol_calibration}. When the forward variance $\widetilde{\xi}$ was modelled with a term structure, the interval $[0.01, 0.16]$ was used for every point on the term structure. The ranges for maturities and strikes were chosen as wide as possible before the pricing and implied volatility routines start returning meaningless results due to numerical noise; this happened more often at short maturities and deep in the money.

The accuracy of the Chebyshev Tensors built was measured by comparing the $1,000$ benchmark (synthetically generated) volatility surfaces with the corresponding surfaces obtained by the Chebyshev Tensors. That is, evaluating the Chebyshev Tensors on the $1,000$ randomly generated parameters, along with the maturities and strikes needed to generate volatility surfaces, and comparing these to the corresponding benchmark surfaces. For each point of the volatility surface, the mean and the maximum errors were computed.


To assess the accuracy of the Chebyshev Tensors in the calibration of the rough Bergomi model, we do the following.\footnote{Calibrations were done using Python and Matlab.} We have $1,000$ volatility surfaces that have been generated synthetically. For each surface, a calibration of the rough Bergomi model can be done by replacing the pricing function $\phi$ (such as the one in Equation \ref{eq: implied vol function price followed by bs}), by the Chebyshev Tensor that approximates $\phi$. The accuracy of each calibration is measured using the root-mean-squared error (RMSE), defined in Equation \ref{eq: root mean squared error}.

\begin{equation}\label{eq: root mean squared error}
\textrm{RMSE} = \sqrt{\dfrac{1}{NM}\sum_{i=1}^N\sum_{j=1}^M(q_{ij} - v_{ij})^2}.
\end{equation}

In Equation \ref{eq: root mean squared error}, the index $i$ runs through maturities, $j$ through strikes, $q_{ij}$ is the synthetically generated implied volatility corresponding to the $i$-th maturity and $j$-th strike, and $v_{ij}$ is the corresponding implied volatility generated by the Chebyshev Tensor.

In addition to accuracy at the level of proxy pricer and calibration, the speed of the tensors was also measured to assess the computational gains obtained during calibration when Chebyshev Tensors are used, compared to calibrations done with the pricing routines of the rough Bergomi model.


The next two Sections present the results obtained for each of the cases of the rough Bergomi model considered. They were obtained using a standard PC computer, with $i7$ cores; no parallelisation nor GPUs were used.

\subsection{Rough Bergomi model with constant forward variance}





As mentioned in the previous Section, the rough Bergomi model with constant forward variance has $4$ parameters and hence Chebyshev Tensors were built to approximate it in two different ways. First, by evaluating each grid point. Then, by running the Rank Adaptive Algorithm to obtain Chebyshev Tensors in TT format. Both tensors had dimension $6$, resulting from $4$ parameters in $\theta$, plus maturity and strike.



\subsubsection{Building cost}

For the Chebyshev Tensors built in the direct way, a mesh of $14,400$ points was used. This was obtained by using the following number of Chebyshev points per dimension $5, 5, 3, 4, 6, 8$. The decision of how many points to use per dimension was made based on one-dimensional cross sectional slices of the pricing function; variables with respect to which the pricer showed greater sensitivity got higher numbers of Chebyshev points; those with less sensitivity got lower. The memory footprint of this Chebyshev Tensor was $800$ KB.

Although these many Chebyshev points translates into $14,400$ calls to the pricing routine, one can reduce this to only $300$ pricing calls. In the Chebyshev mesh mentioned above, there are $300$ parameter combinations in $\Theta$, coming from the first four dimensions, where the number of grid points are $5,5,3,4$. For each of these combinations, a Monte Carlo diffusion of the option underlying can be used to obtain all prices associated with the maturity and strike combinations given by the grid on the last two dimensions, which correspond to maturity and strike.\footnote{This is the way implied volatilities were obtained for a whole surface in \cite{Horvath_vol_calibration}, given a $\Theta$ parameter combination.}


The $300$ parameter combinations ($\Theta$ parameters) mentioned above represent a reduction of $99.62\%$ compared to the reported $80,000$ parameter combinations used in \cite{Horvath_vol_calibration}; 
a reduction of $99.25\%$ compared to the $40,000$ parameter combinations reported in \cite{Horvath_bayer_vol_calib}; and a reduction of $99.97\%$ with respect to the $1,000,000$ evaluations --- combination of parameters, maturities and strikes --- used in \cite{bayer}. 


To build the TT-format Chebyshev Tensor using the Rank Adaptive Algorithm, $7$ Chebyshev points per dimension were used. This corresponds to a grid of $117,649$ points. Notice that the number of grid points is significantly higher than for the Direct Chebyshev Tensor. One can afford this increase in number of grid points, as the Rank Adaptive Algorithm allows for a much lower number of calls to the pricing routine. Therefore, one can increase the granularity of the mesh and, with it, potentially the accuracy of approximation. The function was evaluated on $10,000$ randomly selected grid points, $8.5\%$ of the total grid points. From the $10,000$ grid points evaluated, $8,000$ were used for training and $2,000$ for testing. 

Given that the $10,000$ grid points evaluated are randomly selected, there is no way of reducing the computational cost as in the case of the Direct Chebyshev Tensor, and $10,000$ calls to the pricing function must be made. This represents a computational building cost reduction of $87.5\%$ compared to the $80,000$ parameter combinations in \cite{Horvath_vol_calibration}; a reduction of $75\%$ compared to the $40,000$ parameter combinations reported in \cite{Horvath_bayer_vol_calib}; and a reduction of $99\%$ with respect to the $1,000,000$ evaluations used in \cite{bayer}.

Once these $10,000$ values were obtained, it took the Rank Adaptive Algorithm under $5$ minutes to obtain a TT-format Chebyshev Tensor --- in Matlab. The maximum rank of the TT-format Chebyshev Tensor was $12$ and its memory footprint $20$ KB.

It is important to note that the building cost corresponds to the offline part of the process. The building of Chebyshev Tensor can be done at any point and if needed, in a separate module to the risk system. Moreover, once the Chebyshev Tensor has been built, it can be used in as many calibration exercises as required. In principle, the tensor built should work well for many weeks or monthts in the future. Perhaps one should consider building a new one once market conditions change substantially. Other than that, the object is built once, stored in memory --- something that can easily be done given its low memory footprint --- and loaded to be used whenever needed.


\subsubsection{Accuracy of proxy pricer}

\begin{figure}[H]
\centering
\includegraphics[scale=0.37]{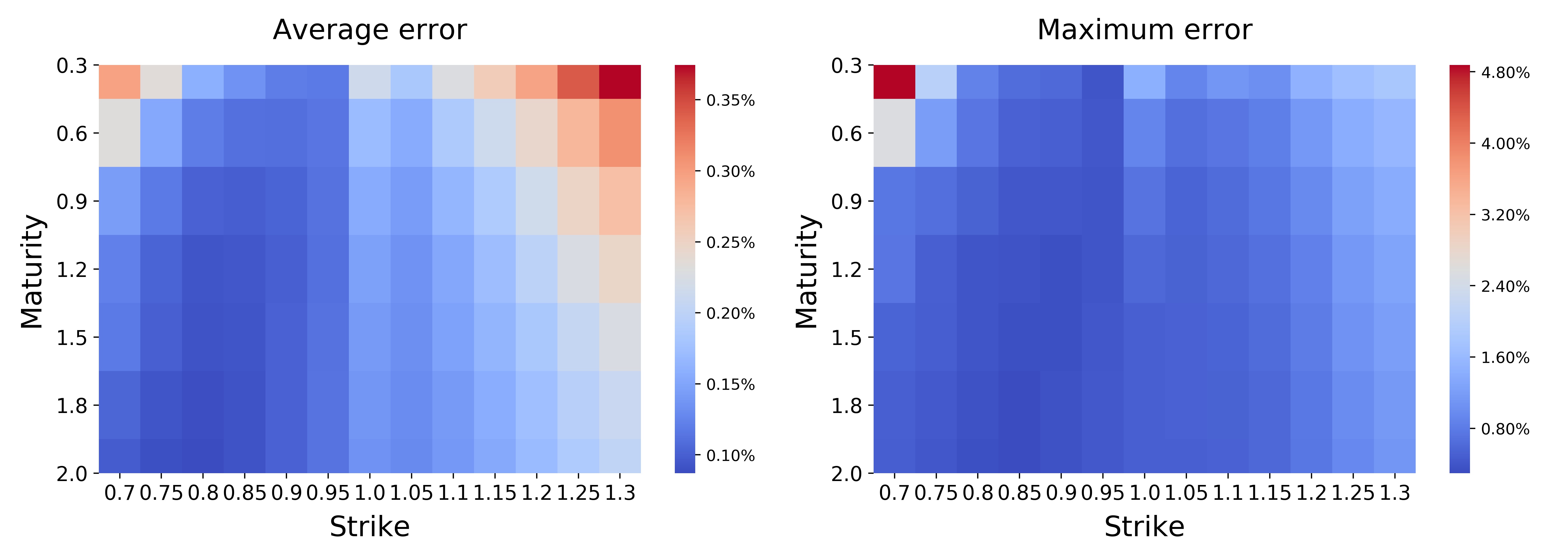}
\caption{Average and maximum error of approximation heat map for Direct Chebyshev Tensors for the rough Bergomi model. It assumes constant forward variance.}
\label{fig: accuracy constant fwd variance cheb tensor naive}
\end{figure}

\begin{figure}[H]
\centering
\includegraphics[scale=0.37]{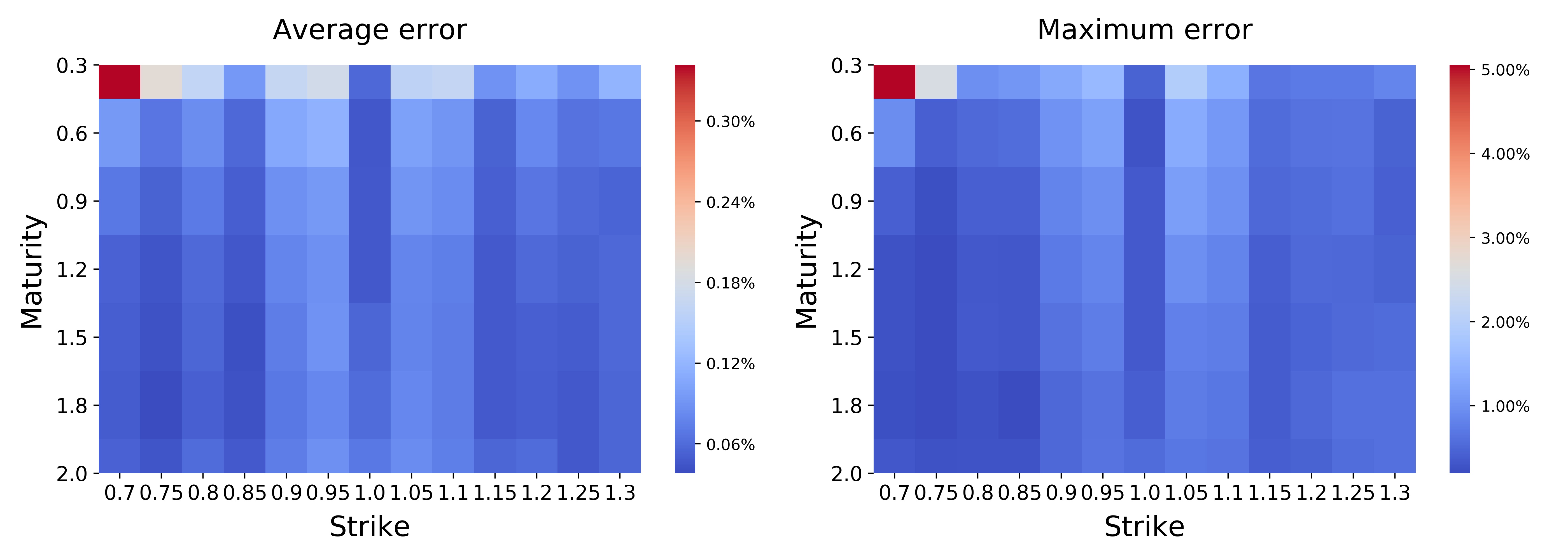}
\caption{Average and maximum error of approximation heat map for TT-format Chebyshev Tensors for the rough Bergomi model. It assumes constant forward variance.}
\label{fig: accuracy constant fwd variance cheb tensor rank adap}
\end{figure}

Figures \ref{fig: accuracy constant fwd variance cheb tensor naive} and \ref{fig: accuracy constant fwd variance cheb tensor rank adap}, show a heat map of the average and maximum error of the proxy pricer, for both proxy methods (Direct Chebyshev Tensor and Chebyshev Tensor in TT-format), across the $1,000$ volatility surfaces. Both methods yield similar results; the average errors fall within $0.05\%$ and $0.35\%$, while the maximum errors are between $0.1\%$ and $5\%$, in absolute implied volatility values.

The accuracy shown in these Figures compares well with the ones presented in \cite{Horvath_vol_calibration} and \cite{bayer}, where Deep Neural Nets were used.

\subsubsection{Accuracy of calibration}

\begin{figure}[H]
\centering
\includegraphics[scale=0.33]{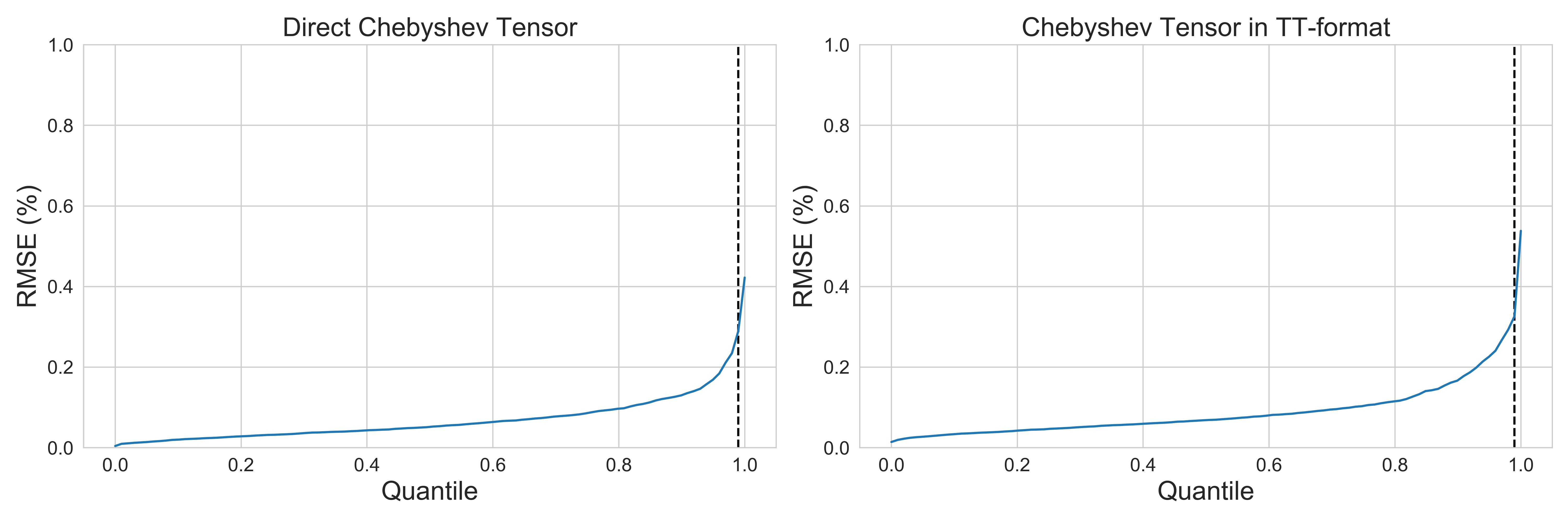}
\caption{Quantiles for the distribution of RMSE values obtained by calibrating $1,000$ synthetically generated implied volatility surfaces using Chebyshev Tensors. The left pane corresponds to the Direct Chebyshev Tensor; the right pane corresponds to the Chebyshev Tensor in TT-format . The vertical line on both plots corresponds to the $99\%$ quantile.}
\label{fig: rmse quantiles flat fwd variance}
\end{figure}

Figure \ref{fig: rmse quantiles flat fwd variance} shows quantile values of the error distribution measured using the root mean square error (RMSE), between the $1,000$ synthetically generated implied volatility surfaces and the $1,000$ implied volatility surfaces obtained after calibrating with the two types of Chebyshev Tensors. Note that in both cases, the $99\%$ quantile of the RMSE is around $0.5\%$ and the maximum RMSE value is less than $0.6\%$.


This accuracy compares very well with the results presented in \cite{Horvath_vol_calibration} and \cite{Horvath_bayer_vol_calib}, where Deep Neural Nets were used. 

\subsubsection{Computational gains at calibration}

Benchmarking the evaluation speed of Chebyshev Tensors is not straightforward given the set up used. The rough Bergomi model is implemented in C++. The Direct Chebyshev Tensors can be built and run in either C++ or Python. The implementation of the Rank Adaptive Algorithm is in Matlab. The is the only implementation known to the authors. The optimisation routines used for the calibration of implied volatility surfaces were taken from Python (in the case where the Direct Chebyshev Tensor is used), and from Matlab (when the TT-format Chebyshev Tensor is used).\footnote{The SLSQP algorithm available in SciPy for Python, and the Least Squares optimisation algorithm in Matlab. In both cases, the algorithms admit domain constraints, which is important for Chebyshev Tensors (see Section \ref{sec: calibration with cheb tensors} for discussion on this point).}

The average evaluation time for the pricing function --- in C++, using $60,000$ paths --- was of $2.6$ seconds. The Direct Chebyshev Tensors had an average evaluation time of $60.5$ microseconds in C++; hence $40,000$ times faster. The evaluation time for the TT-format Chebyshev Tensors, in Matlab, was of $1$ milliseconds. If one assumes between $10$ and $100$ times speed acceleration from Matlab to C++, one would expect to have $10-100$ microseconds in a C++ implementation; hence, a similar evaluation time to that of the Direct Chebyshev Tensors. As a result, the computational gain of the Rough Bergomi model calibration via Chebyshev Tensors is around $40,000$ times faster compared to the one that uses the original Pricing function.



A typical volatility calibration exercise used around $100$ parameter trials. This translates --- assuming $100$ points on the volatility surface --- into $10,000$ calls to the pricer or its proxy. Given the reported time of around $60.5$ microseconds per Chebyshev Tensor evaluation for a C++ object, this means $0.6$ seconds spent calling the tensor during calibration. This massively contrasts with the $7.2$ hours that it would take, should the original pricing function be used.

\subsection{Rough Bergomi model with piecewise-constant forward variance}

The rough Bergomi model with piecewise constant forward variance has $11$ parameters. Once maturity and strike are added to the domain of approximation, the Chebyshev Tensor has dimension $13$. It is impossible to build such a tensor through the direct method given its dimension; the Rank Adaptive Algorithm has to be used and the tensors have to be expressed in TT-format.

\subsubsection{Building cost}

A grid of $7$ points per dimension was chosen for the Chebyshev Tensor. This gives a total of $96,889,010,407$ grid points. Out of these, only $20,000$ were randomly chosen to be evaluated and used in the Rank Adaptive Algorithm. This represents only $\num{2e-5}\%$ of the total grid. From the $20,000$ points evaluated, $16,000$ were used for training, and $4,000$ for testing. It took the Rank Adaptive Algorithm under $8$ minutes to obtain a TT-format Chebyshev Tensor --- in Matlab. The maximum rank of the TT-format Chebyshev Tensor was $9$ and its memory footprint $16$ KB.

Once again, given that we are dealing with $20,000$ randomly selected grid points, we are forced to call the pricing routine on all $20,000$. This, however, still represents a reduction in building cost of $75\%$, $50\%$ and $98\%$, with respect to \cite{Horvath_vol_calibration}, \cite{Horvath_bayer_vol_calib} and \cite{bayer}, respectively.


\subsubsection{Accuracy of the proxy pricer}

\begin{figure}[H]
\centering
\includegraphics[scale=0.37]{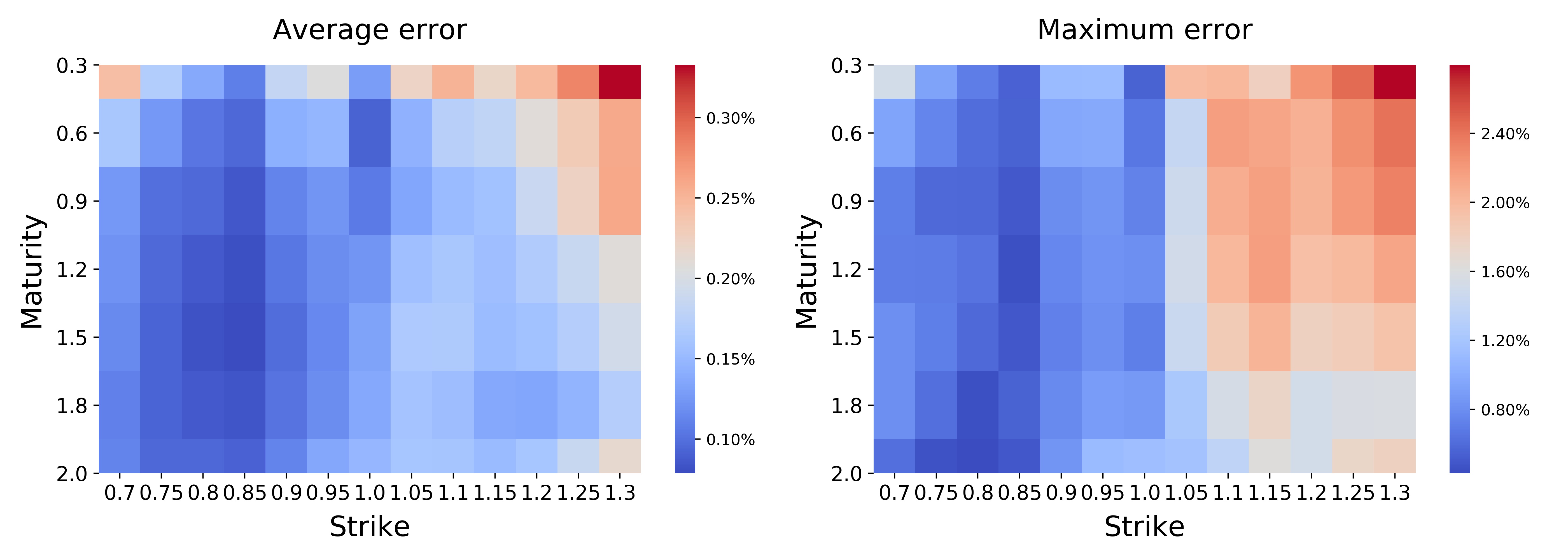}
\caption{Average and maximum error of approximation heat map for Chebyshev Tensors built using the Rank Adaptive Algorithm for the rough Bergomi model. It assumes piecewise forward variance.}
\label{fig: accuracy piecewise fwd variance cheb tensor}
\end{figure}

Figure \ref{fig: accuracy piecewise fwd variance cheb tensor} shows the accuracy of approximation of the built Chebyshev Tensor. 
Once again, accuracy is very high; similar to the one presented in \cite{Horvath_vol_calibration} and \cite{bayer}, where Deep Neural Nets were used.

\subsubsection{Accuracy of calibration}

\begin{figure}[H]
\centering
\includegraphics[scale=0.3]{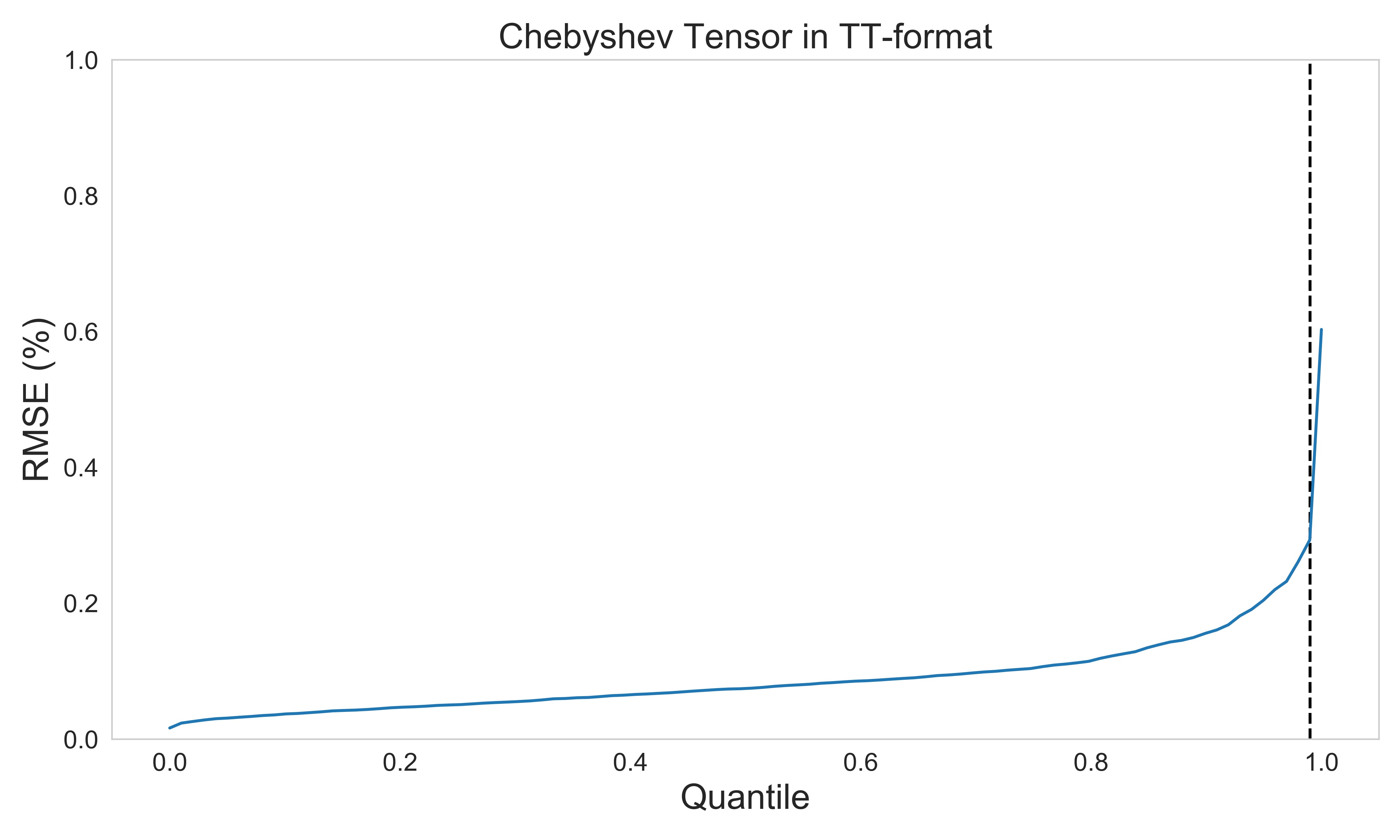}
\caption{Distribution of RMSE values obtained by calibrating $1,000$ synthetically generated implied volatility surfaces using Chebyshev Tensors. The plot corresponds to the TT-format Chebyshev Tensor. The vertical line on both plots corresponds to the $99\%$ quantile.}
\label{fig: rmse quantiles piecewise fwd variance 1}
\end{figure}

The same $1,000$ synthetic implied volatility surfaces were calibrated using this Chebyshev TT-tensor. The quantile values of the error distribution, measured using the RMSE, is shown in Figure \ref{fig: rmse quantiles piecewise fwd variance 1}. It can be appreciated that the maximum error is below $0.6\%$.

Once again, the results compare to the ones presented in \cite{Horvath_vol_calibration} and \cite{bayer}.

\subsubsection{Computational gains at calibration}

The evaluation time of the TT-format Chebyshev Tensor was measured at $2$ milliseconds per evaluation (in Matlab). An implementation in C++ would be expected to run between $10$ and $100$ times faster. Therefore, the calibration routing using the Chebyshev Tensor should be between $10,000-100,000$ times faster than one using the original pricing function. Again, computational gains range from several hours to a few seconds.


\section{Conclusions}\label{sec: conclusion}

Pricing models --- such as the rough Bergomi model in \cite{rough_bayer} --- cannot be used for calibration in practical settings due to the high computational cost involved. This paper gives evidence that by approximating the composition of the rough Bergomi model with the Black-Scholes implied volatility function, using Chebyshev Tensors, calibration of such models becomes feasible in practical settings.

The domain of approximation included all model parameters, along with maturity and strike. The advantage of this approach is that any point of the implied volatility surface can be evaluated. Two cases of the Bergomi model were considered: constant forward variance and piecewise-constant forward variance. Depending on the dimension of the model, Chebyshev Tensors were built either by directly evaluating every grid point (Direct Chebyshev Tensors), or through the use of the Rank Adaptive Algorithm, bypassing the curse of dimensionality (TT-format Chebyshev Tensors).

The accuracy of the Chebyshev Tensors was measured on $1,000$ implied volatility surfaces randomly generated by the pricing model. As can be seen in Figures \ref{fig: accuracy constant fwd variance cheb tensor naive}, \ref{fig: accuracy constant fwd variance cheb tensor rank adap} and \ref{fig: accuracy piecewise fwd variance cheb tensor}, the accuracy is very high; comparable to the ones presented in \cite{Horvath_vol_calibration} and \cite{bayer}, obtained with Deep Neural Nets. However, the number of calls to the pricing function needed to build Chebyshev Tensors is much smaller then the ones reported in \cite{Horvath_vol_calibration} and \cite{bayer}; in some cases $99\%$ more efficient.

Calibration was tested on the same $1,000$ synthetically generated implied volatility surfaces. The quantile values of the errors, measured using the RMSE, are shown in Figures \ref{fig: rmse quantiles flat fwd variance} and \ref{fig: rmse quantiles piecewise fwd variance 1}, where all errors were less than $0.6\%$. 

The speed of evaluation for the Chebyshev Tensors built is such, that a single calibration, with a C++ implementation, should be completed in less than a second; around $40,000$ times faster than if the original pricing function is used.

Chebyshev Tensors are light in memory and can be serialised. Their memory footprint being less than 1 MB; in some case as low as a few KB. This makes them ideal objects for the daily calibration of pricing models, as they can be stored and loaded whenever needed.

The results presented focused on the rough Bergomi model. However, there is no reason why these should not generalise to other models. The approach used makes little to no assumptions specific to the model used in the simulations.

A possible limitation of the approach relates to the dimension of the model parameter space. The greater the dimension, the less effective it is. The maximum dimension tested was $13$. We do not know what will be the maximum dimension for which the presented technique would work. This is left for future research.

Another limitation comes in the choice of domain. Chebyshev Tensors do not extrapolate. Therefore, one is restricted to either optimisation routines that admit domain constraints --- such as the ones considered in this paper --- or calibrations that use a combination of Chebyshev Tensors and pricing model. As mentioned in Section \ref{sec: calibration with cheb tensors}, choosing a large enough domain should solve this problem in most cases. 
The calibration errors presented for the rough Bergomi model show that the domain ranges considered were sufficient to obtain very good results.


\begin{thebibliography}{1}



\bibitem{rough_vol_first_paper} Al\`{o}s, E., Le\'{o}n, J. A., Vives, J. 2007. On the short-time behavior of the implied volatility for jump-diffusion models with stochastic volatility. \emph{Finance and Stochastics} 11 (4) 571-589. 


\bibitem{bayer} Bayer, C. Stemper, B. 2018. Deep calibration of rough stochastic volatility models. \emph{https://arxiv.org/abs/1810.03399}


\bibitem{rough_bayer} Bayer, C. Friz, P. Gatheral, J. 2015. Pricing under rough volatility. \emph{Quantitative Finance} 16(6). 1-18. 




\bibitem{bergomi} Bergomi, L. 2015 \emph{Stochastic Volatility Modeling}. Chapman \& Hall/CRC.


\bibitem{Doleans} Dol\'{e}ans-Dade, C. 1970. Quelques applications de la formule de changement de variables pour les semimartingales. \emph{Wahrscheinlichkeitstheorie verwandte Gebiete} 16. 181-194. 



\bibitem{Gatheral} Gatheral, J. 2011 \emph{The volatility surface: a practitioner's guide}. Wiley.


\bibitem{gatheral_rough_paper} Gatheral, J. Jaisson, T. Rosenbaum, M 2018. Volatility is rough. \emph{Quantitative Finance} 18(6). 933-949. 



\bibitem{GlauParamOptPric} Gaß, M. Glau, K. Mahlstedt, M. and Mair, M. 2018. Chebyshev Interpolation for Parametric Option Pricing. \emph{Finance and Stochastics} 22. 701-731. 


\bibitem{glau_completion} Glau, K. Kressner, F. Statti, F. 2019. Low-rank tensor approximation for Chebyshev interpolation in parametric option pricing. \emph{https://arxiv.org/abs/1902.04367}


\bibitem{SABR} Hagan, P. S., Kumar, D., Lesniewski, A. S., Woodward, D. E. 2002. Managing smile risk. \emph{The Best of Wilmott} 1. 249-296. 


\bibitem{Heston}  Heston, S. L. 1993. A closed-form solution for options with stochastic volatility with applications to bond and currency options. \emph{The Review of Financial Studies} 6 (2). 327-343. 


\bibitem{Andres} Hernandez, A. 2017. Model calibration with neural networks. \emph{Risk}



\bibitem{Horvath_vol_calibration} Horvath, B., Muguruza, A. Tomas, M. 2020. Deep Learning Volatility. \emph{https://arxiv.org/abs/1901.09647}


\bibitem{Horvath_bayer_vol_calib} Bayer, C. Horvath, B., Muguruza, A. Stemper, B. Tomas, M. 2020. On deep calibration of (rough) stochastic volatility models. \emph{https://arxiv.org/abs/1908.08806}


\bibitem{Horvath_rough_vol} Horvath, B., Jacquier, A. Muguruza, A. 2017. Functional central limit theorems for rough volatility. \emph{arXiv:1711.03078, 2017.}



\bibitem{Completion_Steinlechner} Steinlechner, M. Riemannian optimization for high-dimensional tensor completion. \emph{SIAM J. Sci. Comput.} 38.  (2016). S461–S484. 


\bibitem{TT_tensor_Oseledets} Oseledets, I.V. 2011. Tensor-train decomposition. \emph{SIAM J. Sci. Comput.} 33. 2295–2317. 


\bibitem{Runge} Runge, C. 1901. Über empirische Funktionen and die Interpolation zwischen äquidistanten Ordinaten. \emph{Z. Math. Phys.} 224–243.


\bibitem{TrefethenTextbook} Trefethen, L. 2013 \emph{Approximation Theory and Approximation Practice}. SIAM.


\bibitem{Clenshaw_stability} Smoktunowicz, A. 2002. Backward Stability of Clenshaw's Algorithm. \emph{BIT Numerical Mathematics} 42. 600-610. 


\bibitem{cheb_slider} Zeron, M. Ruiz, I. 2019. Denting the FRTB IMA computational challenge via Orthogonal Chebyshev Sliding Technique. \emph{arxiv.org/abs/1911.10948}




\end{thebibliography}
\end{document}